\documentclass[12pt,preprint]{aastex}

\shorttitle{The initial spikes of SGR giant flares}
\shortauthors{Tanaka et al.}

\begin{document}


%\title{The initial spike of the 1998 August 27 giant flare from SGR 1900+14:\\
% A comparative study of \textit{GEOTAIL} observations}
\title{Comparative study of the inial spikes of SGR giant flares\\
in 1998 and 2004 observed with GEOTAIL:\\
Do magnetospheric instabilities trigger large scale fracturing 
of magnetar's crust?}


\author{Y. T. Tanaka\altaffilmark{1}, T. Terasawa\altaffilmark{2}, 
N. Kawai\altaffilmark{2}, A. Yoshida\altaffilmark{3},
I. Yoshikawa\altaffilmark{1}, Y. Saito\altaffilmark{4},\\
T. Takashima\altaffilmark{4}, and T. Mukai\altaffilmark{4}  }
%\affil{Department of Earth and Planetary Science, University of Tokyo, Japan}
\email{yasuyuki@eps.s.u-tokyo.ac.jp}

%\author{T. Terasawa\altaffilmark{2}, N. Kawai\altaffilmark{2}}
%\affil{Department of Physics, 
%Tokyo Institute of Technology, Japan}

%\author{A. Yoshida\altaffilmark{3}}
%\affil{Department of Physics and Mathematics, 
%	Aoyama Gakuin University, Japan}

%\author{I. Yoshikawa\altaffilmark{1}, K. Yoshioka\altaffilmark{1}}
%\affil{Department of Earth and Planetary Science, University of Tokyo, Japan}

%\and

%\author{Y. Saito\altaffilmark{4}, T. Takashima\altaffilmark{4}, T. Mukai\altaffilmark{4}}
%\affil{Japan Aerospace Exploration Agency, Japan}


\altaffiltext{1}{Department of Earth and Planetary Science, 
University of Tokyo, Japan}
\altaffiltext{2}{Department of Physics, Tokyo Institute of Technology, Japan}
\altaffiltext{3}{Department of Physics and Mathematics, Aoyama Gakuin University, Japan}
\altaffiltext{4}{Japan Aerospace Exploration Agency, Japan}
%\altaffiltext{5}{Patron, Alonso's Bar and Grill}


\begin{abstract}
We present the unsaturated peak profile of SGR 1900+14
giant flare on 1998 August 27. This was obtained by 
particle counters of the Low Energy Particle instrument
onboard the GEOTAIL spacecraft.
The observed peak profile revealed four characteristic structures: 
initial steep rise, intermediate rise to the peak, exponential decay 
and small hump in the decay phase. 
From this light curve, we found that the isotropic peak luminosity 
was $2.3\times10^{46}$ erg s$^{-1}$ and the total energy was 
$4.3 \times 10^{44}$ erg s$^{-1}$ ($E\gtrsim$ 50 keV),
assuming that the distance to SGR 1900+14 is 15 kpc and that the spectrum
is optically thin thermal bremsstrahlung with $kT =$ 240 keV.  
These are consistent with the previously reported lower limits derived from 
Ulysses and Konus-Wind observations. A comparative study of the initial spikes 
of SGR 1900+14 giant flare in 1998 and 
SGR 1806-20 in 2004 is also presented.
The timescale of the initial steep rise shows the magnetospheric origin, 
while the timescale of the intermediate rise to the peak 
indicates that it originates from the crustal fracturing.
Finally, we argue that the four structures and their corresponding
timescales provide a clue to identify extragalactic 
SGR giant flares among short GRBs.
 
\end{abstract}





\section{Introduction}

Soft gamma-ray repeaters (SGRs) were first discovered 
as high-energy burst sources in the late 1970's \citep{Mazets1981}. 
Once SGRs enter burst active phases,
they produce a lot of short-duration ($\sim$0.1 s) energetic ($\sim10^{41}$
erg) soft gamma-ray bursts. These bursts were distinguished 
from cosmological gamma-ray bursts (GRBs) 
by the soft spectra and the repeated activities.
Furthermore, as rare events, SGRs emit extremely bright giant flares
(GFs). A GF lasts for several hundred seconds and 
its isotropic total energy amounts to 10$^{44}-10^{46}$ erg.
%Furthermore, one of the most striking features 
%To date, four SGRs are confirmed to exist.
So far, only three have been recorded. On 1979 March 5,
the first GF was detected from SGR 0526-66 by Venela spacecraft
\citep{Mazets1979}. The second GF was observed 
from SGR 1900+14 on 27 August 1998 \citep{Hurley1999,Mazets1999,Feroci2001}. 
Recently SGR 1806-20 emitted the third GF on 27 December 2004 
\citep{terasawa2005,Hurley2005,Palmer2005,Mereghetti2005,Schwartz2005}.
The overall time profile of each GF is characterized 
by a very intense spectrally hard initial spike 
whose duration is $\lesssim$ 0.5 s, 
and a subsequent pulsating tail which has a softer spectrum and lasts for
some hundred seconds. After the GFs, radio afterglows were observed
from SGR 1900+14 \citep{Frail1999}
and from SGR 1806-20 \citep{Gaensler2005,Cameron2005}.

SGRs show the slow spin periods ($5-8$ s) and 
rapid spin-down rates ($10^{-11}-10^{-10}$ s s$^{-1}$)
\citep{Kouveliotou1998,Kouveliotou1999}.
Assuming magnetic dipole radiation, 
%From observations of slow spin periods ($5-8$ s) and 
%rapid spin-down rates ($10^{-11}-10^{-10}$ s s$^{-1}$),  
we can estimate the magnetic fields of SGRs to be $10^{14}-10^{15}$ G
and SGRs are recognized 
as magnetars \citep{Duncan1992,Thompson1995,Thompson1996}.
According to the magnetar model,
the energy source of both recurrent bursts and GFs is the
ultrastrong magnetic field: 
stored magnetic energy inside a magnetar is suddenly released
via cracking of a magnetar's crust, and the large scale crustal
fracturing produces GFs.
%both recurrent bursts and giant flares result from the crack of 
%the neutron star crust produced by magnetic stresses. 
Similar to earthquakes, the power-law distribution 
of the radiated energy of the repeated burst and the lognormal distribution 
of waiting times between successive bursts are reported \citep{Cheng1996,
Gogus2000}. These observations also support the idea 
that SGR bursts originate from the starquakes.
%The energy source of
%the quiescent X-ray emission of $\sim10^{35}$ erg s$^{-1}$ 
%is also the magnetic field, because
%the rotational energy loss is insufficient by several orders of magnitude.
%The decay of the magnetic field inside the neutron star heats the surface
%and persistent thermal soft X-rays are emitted. 

%It is widely recognized that SGRs are magnetars 
%\citep{Duncan1992,Thompson1995,Thompson1996},
%slowly rotating ($5-8$ s) neutron stars with ultrastrong magnetic fields of
%$10^{14}-10^{15}$ G. 
%Detection of a
%rapid spin down also confirm this theory.
%While this flare is observed by Ulysses and Konus-Wind spacecraft, particle
%detectors onboard GEOTAIL satellite determined the peak profile. 
%We also show the total emitted energy of the giant flare.

In this paper, first, we focus on the SGR 1900+14 GF on 1998 August 27.
This flare was detected by gamma-ray instruments on the
Ulysses, Konus-Winds and BeppoSAX satellites
\citep{Hurley1999,Mazets1999,Feroci2001}. 
However the flare was so intense that these instruments 
underwent severe dead-time or pulse pile-up problems.
Consequently, the time profile during the most intense period
was not obtained and only the
lower limits of the peak flux intensity and fluence were reported
\citep{Hurley1999,Mazets1999}.
Here we present the clear peak profile of the SGR 1900+14 GF on 1998 August 27.
The profile was recorded by the Low Energy Particle instrument (hereafter LEP) 
\citep{Mukai1994} onboard the GEOTAIL spacecraft, whose principal 
objective is to study the Earth's magnetosphere. 
The light curve for the first 350 ms of the GF 
is unsaturated and has a high time resolution 
of 5.58 ms. We also show the energetics of the flare.

Second, we present a comparative study of the initial spikes of 
SGR GFs in 1998 and 2004, the latter of which was also detected by the same
instrument \citep{terasawa2005}. From both of the light curves,
we extract the characteristics of the initial spikes of SGR GFs, 
focusing on the timescales discovered during the initial spikes.
Finally we argue that the observed timescales may provide a clue to
identify extragalactic SGR giant flares among short GRBs.

\section{Instrumentation and Observation}

%The data was drawn from the LEP onboard the GEOTAIL spacecraft.
The LEP is designed to measure three-dimensional velocity distributions
of the Earth's magnetospheric ions and electrons.
%It has a uniform field of view over 140$^{\circ}$ in a meridian plane
%relative to the spacecraft spin axis. Since the spacecraft rotates
%every $\sim$3 s, the LEP covers all azimuthal directions.
%The LEP has been described in detail
%(see \citet{Mukai1994} for the details of the instrument), so we will
%only summarize here its most relevant characteristics.
It consists of two nested sets of quadspherical electrostatic 
analyzers; one analyzer to select ions, and the other to select electrons.
At the receving end of the ion and electron optics,
seven microchannel plate detectors (MCPs) and seven channel electron
multipliers (CEMs) are used, respectively.
During the SGR 1806-20 GF in 2004, the peak flux was so intense that 
the MCPs were saturated during the first 150 ms. 
Alternatively the peak profile was derived from the CEMs, because
the CEMs are much less sensitive to gamma-rays than
the MCPs. After the most intense period, the MCPs recovered from the saturation
and observed the decay profile clearly.
On the other hand, during the SGR 1900+14 GF in 1998, 
we obtained the peak profile from the MCPs. The peak flux of 
the 1998 GF was about one-tenth of that of the 2004 GF (see below), 
and hence the MCPs did not suffer the severe saturation problem.
The CEMs showed count increases ($\lesssim$ 20) corresponding to
those of the MCPs. However, since the background electron counts
for CEMs were high ($\sim$50$-$80), we do not use the CEM data for the
analysis of the SGR 1900+14 GF.
 
%The spacecraft rotates with a period of $\sim$3 s.
%and the spin axis is nearly perpendicular to the elliptic plane.
The LEP records the data every 15/8192 of the spacecraft 
spin period over 32 sequences,
followed by a gap of 1/256 of the spin period. The spacecraft
spin period was 3.046 s on 1998 August 27, leading to 
$3.046 \times \left( 15/8192 \right) = 5.58\times10^{-3}$ s $= 5.58$ ms 
time resolution.
%Gamma-rays from the GF directly stimulate the MCP detectors.
%The LEP record the counts in the following way.
%It azimuth angle 
%This leads to a 5.58 ms time resolution.
This is slightly different compared to a 5.48 ms time resolution 
of SGR 1806-20 GF observation in 2004, during which the spin period
was 2.993 s.

In this report, we use the LEP calibration that the effective energy range
and the detection efficiency are $>\sim$ 50 keV and $\sim$1\% against
incident photons, respectively. Since the LEP was not designed to
measure gamma-rays, this calibration was made after the launch of the GEOTAIL
spacecraft through the analyses of solar flare photons for which
the Hard X-ray Telescope onboard the Yohkoh satellite \citep{Kosugi1991} 
provided photon energy spectra and intensities.
Recently we have made (i) GEANT4 simulations based on the detailed mass model
of the LEP, satellite structure and other instruments, and
(ii) the laboratory measurements
of the detection efficiency of the MCP \citep{Tanaka2007}, both of which have 
successfully reproduced what were obtained from the solar flare
photon analyses. In addition, we found from the GEANT4 simulations that
the effect of the rotation of the spacecraft was negligibly small
around the spin phase angles corresponding to the two GFs. 

%\section{Observation: SGR 1900+14 giant flare on 27 August 1998}

%Gamma-rays from the giant flare from SGR 1900+14 triggered the LEP
%at 10:22:15 UT of 1998 August 27. At that time, GEOTAIL
%was in the earth's magnetosphere and the LEP is measuring the
%magnetospheric ions and electrons routinely.
Fig. 1 shows the first 350 ms unsaturated peak profile 
of the GF from SGR 1900+14 on 27 August 1998. 
%The time resolution is 5.58 ms. 
%The energy of the detected photons is above $\sim$50 keV, 
%which is confirmed by Monte Carlo simulations based on the Geant4.
%The photon energy is above $\sim$50 keV.
Dead time and saturation effects are negligible for the count rates
smaller than $\sim$1000 counts per 5.58 ms: 
only the peak count at $t$=5.58 ms was dead-time corrected.
The shaded bars in Fig. 1 indicate the instrumental data gaps of 12 ms.
The onset time ($t$=0) was 10:22:15.47 UT, which coincided with the
expected arrival time at the GEOTAIL position.
Before the onset, the count was less than 25 counts per 5.58 ms (shown 
by a black arrow in Fig. 1(b)), i.e. the background level.
Then it increased to 792 counts within 5.58 ms, and this rapid increase
provided the upper
limit of the e-folding time of the initial rise as 1.6 ms.  
After the onset, it reached a very sharp peak of 4776 counts 
at $t$=5.58 ms. This increase yielded the e-folding time of 
the intermediate rise time to the peak as $3.1^{+0.9}_{-2.0}$ ms.
Following the peak, it decayed rapidly. The exponential decay time was
calculated as 2.9$\pm$0.2 ms from the counts for $t$=5.6$-$22 ms. 
Note that the timing of the dip at $t$=22 ms corresponds to 
the timing of the temporal count recovery from the total shut down
of the Konus-Wind instrument (see Fig. 6 of \citet{Mazets1999}). 
After that, it increased again with e-folding time of 16$\pm$2.5 ms 
for $t$=22$-$50 ms and reached a flat-top second peak during 60$-$120 ms.
%Then, the photon counts again increased and reached the flat-top second peak 
%during 60$-$120 ms. 
Finally the exponential decay was clearly observed and the decay
time was obtained as 23$\pm$1.6 ms during $t$=120$-$160 ms.
Note that a small hump was seen around 310 ms, which was
also observed with the Konus-Wind instrument (Fig.6 of \citet{Mazets1999}).

To convert physical quantities such as an energy flux 
from the observed count rates, we need an
assumption on the photon energy spectrum, because the LEP detected 
integrated photon numbers above 50 keV. We assume $kT$=240 keV 
optically thin thermal bremsstrahlung (OTTB) spectrum which was
obtained from Ulysses observation \citep{Hurley1999}.
%and (ii) $kT$=230 keV Blackbody spectrum which was observed during
%the initial spike of the SGR 1806-20 GF in 2004 \citep{Boggs2006}.
Resultant physical quantities are tabulated in Table 1, combined with 
Venela observation of the SGR 0526-66 GF in 1979 \citep{Mazets1999}
and GEOTAIL observation of the SGR 1806-20 GF in 2004 \citep{terasawa2005}.
We found that the peak luminosity and the total emitted energy 
are $2.3\times10^{46} d^2_{15}$ erg s$^{-1}$ and 
$4.3\times10^{44} d^2_{15}$ erg ($E \gtrsim$ 50 keV), respectively. 
Here we assume that the distance to SGR 1900+14 is 15 kpc 
\citep{Vrba2000} and $d_{15}=\left( d/15 \mathrm{kpc} \right)$.
%If we take the blackbody spectrum, the peak luminosity is
%$6.4\times10^{46} d^2_{15}$ erg s$^{-1}$
%and the total emitted energy is $1.2\times10^{45} d^2_{15}$ erg.
%In Table 2, we compare the energetics of the initial spikes of 
%the ever recorded three GFs, combined with 
%Venela observation of the SGR 0526-66 GF in 1979 \citep{Mazets1999}
%and GEOTAIL observation of the SGR 1806-20 GF in 2004 \citep{terasawa2005}.
We also found that the total energy of this GF is about 130 times smaller than
that of the 2004 December 27 GF from SGR 1806-20,
although it is reported that the energy emitted during the pusating tail 
in each GF is comparable ($E_{\rm tail} \sim 10^{44}$ erg, see Table 1).
\citep{Hurley2005,Palmer2005,Mazets1999}. Note that this difference
by a factor of 130 is the same order of the radio observations: 
the radio afterglow of the SGR 1900+14 GF is approximately 500 times 
fainter than that of the SGR 1806-20 GF 
\citep{Frail1999,Gaensler2005,Cameron2005}.

%Is the dip during $t$=15$-$45 ms real? 
%By comparing the time profile observed by the konus-Wind satellite,
%We confirm that the dip is real.

%% In a manner similar to \objectname authors can provide links to dataset
%% hosted at participating data centers via the \dataset{} command.  The
%% second curly bracket argument is printed in the text while the first
%% parentheses argument serves as the valid data set identifier.  Large
%% lists of data set are best provided in a table (see Table 3 for an example).
%% Valid data set identifiers should be obtained from the data center that
%% is currently hosting the data.
%%
%% Note that AASTeX interprets everything between the curly braces in the 
%% macro as regular text, so any special characters, e.g. "#" or "_," must be 
%% preceded by a backslash. Otherwise, you will get a LaTeX error when you 
%% compile your manuscript.  Special characters do not 
%% need to be escaped in the optional, square-bracket argument.

%% In this section, we use  the \subsection command to set off
%% a subsection.  \footnote is used to insert a footnote to the text.


%% This section also includes several of the displayed math environments
%% mentioned in the Author Guide.

\section{Discussion}
%\subsection{Blackbody radius of the initial spike of the SGR 1900+14 GF}
%The spectrum observed during the initial spike of the SGR 1900+14 GF
%was optically thin thermal bremsstrahlung with $kT$ = 240 keV 
%\citep{Hurley1999}. However, due to the similarity of the spectral shape,
%here we assume $kT$= 80 keV blackbody spectrum.
%The temperature is determined as 80 keV, 
%since the blackbody spectrum reaches its peak at $\sim$ 3$kT$ and
%above which it decays similarly to OTTB.
%We define the averaged luminosity during the initial spike $L_{\rm ave}$
%by dividing the total energy
%$E_{\rm total} = 4.3 \times 10^{44}$ erg by the duration of $\sim$0.2 s.
%Then we derive the blackbody radius 
%$R=\left( L_{\rm ave}/4 \pi \sigma T^4 \right)^{1/2}$ = 20 km, 
%nearly coincides with the magnetar's radius. 
%Note that the blackbody radius of 18 km is obtained 
%from the main peak of the SGR 1806-20 GF \citep{Boggs2006}.

%\subsection{Comparative study of the initial spikes}

We observed two SGR GFs out of ever recorded three: 
from SGR 1900+14 in 1998 and
SGR 1806-20 in 2004. Here we present a comparative study and 
extract characteristics of the initial spikes of the SGR GFs.
Fig. 1 and Fig. 2 show the light curves of the initial spikes of
SGR 1900+14 GF and SGR 1806-20 GF, respectively.
Fig. 3 shows the detailed initial rise profiles of both GFs.
From these light curves, we identify four common features: 
(1) initial steep rise (2) intermediate rise to the peak
(3) exponential decay (4) small hump in the decay phase. 
The calculated e-folding times corresponding to the structures 
of (1)-(3) and the timing when we observed the structure (4) 
are tabulated in Table 1. 
%The observed timescales are tabulated in Table 1.
%, combined with the Konus observation of SGR 0526-66. 

First, we focus on (1) initial steep rise. The observed initial rise time 
of SGR 1900+14 GF is $\le1.6$ ms. This is comparable to the
initial rise time of $\le1.3$ ms observed in the SGR 1806-20 GF,
implying the same physical mechanism producing the initial rapid
energy release of these two GFs. Note that
in the leading edge of the initial spike of SGR 1806-20 GF,
Swift and Rhessi observed the similar timescale
(Swift: $\sim$ 0.3 ms, Rhessi: $0.38\pm0.04$ ms) \citep{Palmer2005,Boggs2006}. 
These correspond to our observation of $\le1.3$ ms initial rise time. 
%These time scale corresponds
%to SWIFT and RHESSI observations.
%Note that SWIFT and RHESSI also observed the similar timescale 
%in the leading edge of the initial spike of SGR 1806-20 giant flare
%(SWIFT: $\sim$ 0.3 ms, RHESSI: $0.38\pm0.04$ ms)
%\citep{Palmer2005,Boggs2006}. 
%This similarity meaning the same physical
%mechanism producing the rapid energy release.
%Note that similar timescale was also observed in the SGR 1806-20 by RHESSI
%and SWIFT \citep{Hurley2005,Palmer2005,Boggs2006}.
%The initial rise time of $\le1.5$ ms corresonds to the timescale
%observed with RHESSI and SWIFT \citep{Hurley2005,Palmer2005,Boggs2006}.
According to 
the reconnection model of GFs \citep{Thompson1995,Duncan2004},
%the Alfven crossing time within the magnetosphere of a magnetar 
%$\tau_{\rm mag}$ is comparable to the light-crossing time of the star
%$\tau_{\rm mag} \sim R/c \sim 0.03$ ms, where $R\sim10$ km and 
%$c$ is the speed of light. However, 
reconnection typically occurs 
at a fraction of the Alfven velocity \citep{Thompson1995,Duncan2004},
and this interpretation leads to
$\tau_{\rm mag} \sim L/0.1V_{\rm A} \sim 0.3 \left( L / 10 \rm km \right)$ ms,
where $L$ is the scale of the reconnection-unstable zone, and 
$V_{\rm A} \sim c$ is the Alfven velocity in the magnetosphere. 
%This timescale corresponds to the Alfven crossing time within the
%magnetosphere of a magnetar $\tau_{\rm grow} \sim R/0.1V_{\rm A} \sim 0.3$ ms,
%where $R\sim10$ km and $V_{\rm A} \sim c$ is the Alfven velocity 
%in the magnetosphere and $c$ is the speed of light. 
This theoretical timescale $\tau_{\rm mag}$ seems consistent with the
observation of the initial rise time.

Next, we consider (2) intermediate rise to the peak. 
The observed e-folding rise time of the SGR 1900+14
GF is 3.1 ms, which is shorter than the 9.4 ms rise time 
observed in the SGR 1806-20 GF by factor of about 3.0. 
If this timescale is limited by the propagation of a fracture,
we can infer the fracture size $l$ as 
$l\sim 4 \mathrm{km} \left( t_{\mathrm{rise}}/
4 \mathrm{ms} \right)$ \citep{Thompson2001}.
%we can estimate the fracture size on the basis of this timescale. 
Using this, the fracture size of the SGR 1900+14  
is estimated as $\sim$ 3.1 km, and that of the SGR 1806-20
is as $\sim9.4$ km. It should be noted that our 9.4 ms rise time
observed in the SGR 1806-20 GF differs by factor of $\sim$2 from
4.9 ms derived from the CLUSTER spacecraft observation of the
same GF \citep{Schwartz2005}. The origin of the difference between
these time scales is not understood, but could possibly attribute
to the different energy coverages of the detectors.
Unfortunately, since the energy response of the CLUSTER detectors
against incoming X-ray and gamma-ray photons was not calibrated,
further quantitative comparison between GEOTAIL and CLUSTER
is not possible.

%We speculate that the total emitted energy $E_{\rm total}$ of each GF 
%is proportional to the magnetic energy stored in the patch size $l^2$
%of the SGR's crust. ($E_{\rm total} \propto B_{\rm crust}^2 l^2$). 
%this difference of the fracture size $l$ 
%as well as the magnetic field strength $B$ reflects 
%the difference of the total emitted energy $E_{\rm total}$.
%Since $E_{\rm total}$ and $l$ differs by a factor of 130 and
%3.0 between the SGR 1900+14 GF and SGR 1806-20 GF, we infer that 
%the magnetic field strength of the crust of the SGR 1806-20 
%is 3.6 times stronger than that of the SGR 1900+14.
%This is slightly different compared to the surface magnetic field strength 
%estimated from observations of the period and the period derivative 
%(SGR 1806-20: $7.8 \times 10^{14}$ G, and SGR 1900+14: $5.7 \times 10^{14}$ G
%\citep{Woods2004}).
%we obtain a relation of $E_{\rm total} \propto l^{4.4}$, 
%whose power-law index significantly differs from the natural assumption 
%that $E_{\rm total}$ is proportional to the fracture size $l^2$
%\citep{Thompson1995,Harding2006}. 

In the initial spike of the SGR 1900+14 GF in 1998, we found a deep dip
and rebrightening following a sharp peak (Fig. 1). 
We propose that this dip explains the temporal recovery of the counter
of the Konus-Wind \citep{Mazets1999}, since the dip and the recovery
occurred nearly simultaneously. 
Note that Swift and Rhessi also detected a dip and rebrightening 
in the leading edge of the initial spike of 
the SGR 1806-20 GF \citep{Palmer2005,Boggs2006}, which could not be resolved
by the GEOTAIL observation.
This association implies that the dip and rebrightening are common features
of the initial spikes of the SGR GFs, although theoretical interpretation 
is unclear. 

Then, we concentrate on (3) exponential decay. 
%The physical mechanism of the exponential decay is unknown so far.
The decay time of the SGR 1900+14 GF is 23 ms.
This is shorter than the 66 ms decay time 
of the SGR 1806-20 GF by factor of 2.9, 
which roughly coincides with the factor 3.0
%while that of the SGR 1806-20 GF is 66 ms. 
found in the intermediate rise times. 
From this similarity, we infer that the decay time is also 
proportional to the fracture size of a magnetar's crust.

Finally, we focus on (4) small hump in the decay phase. Small humps are 
observed nearly at the same timing; $\sim$310 ms in 1998 and $\sim$430 ms
in 2004 (note that the hump in 2004 GF was also observed with Swift
satellite \citep{Palmer2005}), although the total emitted energy 
differs by a factor of 130. 
This implies that the hump is caused by the continuing energy
injections rather than the environmental interactions of the flare ejecta.

To conclude, the observed initial rise times imply that 
the onsets of both of the GFs result from magnetospheric instabilities. 
The intermediate rise times, on the other hand, 
are consistent with the idea that main energy release
mechanism of the GFs is the large scale crustal fracturing.
For this interpretation to be valid, magnetospheric instabilities 
should trigger the cracking of a magnetar's crust.
Further theoretical study is needed.

%\subsection{SGR giant flares and short GRBs}

The above four structures discovered in the initial spikes
may provide a clue to identify extragalactic SGR GFs among short GRBs. 
Recently, a possible detection of an extragalactic SGR GF
is reported \citep{2005GCN}. Bright short GRB 051103 was localized
%Recently the bright short GRB 051103 was localized 
near the M81/M82 galaxy group by the interplanetary network.
This association implies that the GRB 051103 is the SGR GF
outside the local group. Furthermore, if the GRB 051103 is emitted from
a SGR in M81, the isotropic total energy amounts to 
$\sim 7 \times 10^{46}$ erg, which
is the same order of the energy of SGR 1806-20 GF \citep{Frederiks2006}.
Not only existence of star forming regions
inside the IPN error quadrilateral of GRB 051103
but also no detection of optical and radio afterglow support 
the SGR hypothesis \citep{Ofek2006a}. 
Here we investigate the hypothesis from the viewpoint of its light curve.

(1) The light curve of GRB 051103 observed by Konus-Wind 
showed a steep rise and the timescale is reported as $\leq$ 6 ms
\citep{Frederiks2006}. This nearly corresponds to the  
%This is the same order of magnitude as the
intermediate rise time of a galactic SGR GF presented above,
although we do not know whether the timescale observed by Konus-Wind
represents an initial rise time or an intermediate rise time.
Furthermore,(2) quasi-exponential decay was seen and the
decay time is $\sim$ 55 ms \citep{Frederiks2006}.
This timescale is also the same order of magnitude as 
the decay times presented above. 
These two similarities found in the light curves
also support the SGR hypothesis.
A hump in a decay phase was not seen in the light curve of 
GRB 051103. This is explicable in terms of the detector's 
detection limit, because the flux of the humps, if exists, are expected
to be about one hundredth of the peak flux.

\acknowledgments

We thank R. Yamazaki for valuable comments and discussions. We are also 
grateful to all the members of GEOTAIL team for their collaboration. 
Y.T.T. is receiving a financial support from JSPS.

\clearpage

\begin{table}
\begin{center}
\caption{Comparison of three SGR giant flares}
\begin{tabular}{lrrr}
\tableline\tableline
 & \multicolumn{1}{c}{SGR 1900+14} & \multicolumn{1}{c}{SGR 1806-20} 
& \multicolumn{1}{c}{SGR 0526-66} \\
\tableline
\tableline
\multicolumn{1}{c}{Initial Spike} & & & \\
E-folding initial rise time [ms] & $<1.6$ & $<1.3$ & $<2$ \\
E-folding intermediate rise time [ms] & $3.1^{+0.9}_{-2.0}$ & 9.4$\pm$1.1 & - \\
Exponential decay time [ms] & 23$\pm$1.6 & 66$\pm$12 & $\sim40$ \\
Timing of Small hump [ms] & $\sim$310 & $\sim$430 & - \\
\tableline
 Peak photon flux [photons cm$^{-2}$ s$^{-1}$] & $\left(3.2_{-1.1}^{+4.0}\right)\times10^6$ & $\left(2.5_{-0.6}^{+1.1}\right)\times10^7$ & - \\
 Peak flux [erg cm$^{-2}$ s$^{-1}$] & $0.85_{-0.30}^{+1.0}$  & $19_{-4}^{+9}$  & $1\times10^{-3}$ \\
 Peak luminosity [erg s$^{-1}$] & $\left(2.3_{-0.8}^{+2.7}\right)\times10^{46}d_{15}^2$  & $\left(5.1_{-1.2}^{+2.3}\right)\times10^{47} d_{15}^2$ & $3.6\times10^{44} d_{55}^2$ \\
 Fluence [erg cm$^{-2}$] & $\left(1.6_{-0.6}^{+2.0}\right)\times10^{-2}$ & $2.0_{-0.5}^{+0.9}$ & $4.5\times10^{-4}$ \\
 Total Energy [erg] & $\left(4.3_{-1.5}^{+5.3}\right)\times10^{44} d_{15}^2$  & $\left(5.4_{-1.3}^{+2.4}\right)\times10^{46} d_{15}^2$ & $1.6\times10^{44} d_{55}^2$ \\
Energy range  & $E$$\gtrsim$50 keV & $E$$\gtrsim$50 keV & $E$$>$30 keV\\
\tableline
\tableline
\multicolumn{1}{c}{Pulsating Tail} & & & \\
Tail Energy [erg] & $1.2\times10^{44} d_{15}^2$ & $1.2\times10^{44} d_{15}^2$ & $3.6\times10^{44} d_{55}^2$ \\
Energy range & $E$$>$15 keV& 3$<$$E$$<$100 keV & $E$$>$30 keV\\
\tableline\tableline
\multicolumn{1}{c}{Reference} & 1 & 2, 3 & 1 \\
%\multicolumn{1}{c}{Reference} & \multicolumn{1}{c}{1} 
%& \multicolumn{1}{c}{2, 3} & \multicolumn{1}{c}{1} \\
\tableline
\end{tabular}
%% Any table notes must follow the \end{tabular} command.
\tablerefs{(1) Mazets et al. 1999; (2) Terasawa et al. 2005; 
(3) Hurley et al. 2005}
\end{center}
\end{table}

\clearpage

%% the file, uncomment the \plotone, \plottwo, and
%% If you need a layout that cannot be achieved with \plotone or
%% \plottwo, you can invoke the graphicx package directly with the
%% \includegraphics command or use \plotfiddle. For more information,
%% please see the tutorial on "Using Electronic Art with AASTeX" in the
%% documentation section at the AASTeX Web site,
%% http://www.journals.uchicago.edu/AAS/AASTeX.
%%
%% The examples below also include sample markup for submission of
%% supplemental electronic materials. As always, be sure to check
%% the instructions to authors for the journal you are submitting to
%% for specific submissions guidelines as they vary from

%% This example uses \plotone to include an EPS file scaled to
%% 80% of its natural size with \epsscale. Its caption
%% has been written to indicate that additional figure parts will be
%% available in the electronic journal.

\begin{figure}
\epsscale{.8}
\plotone{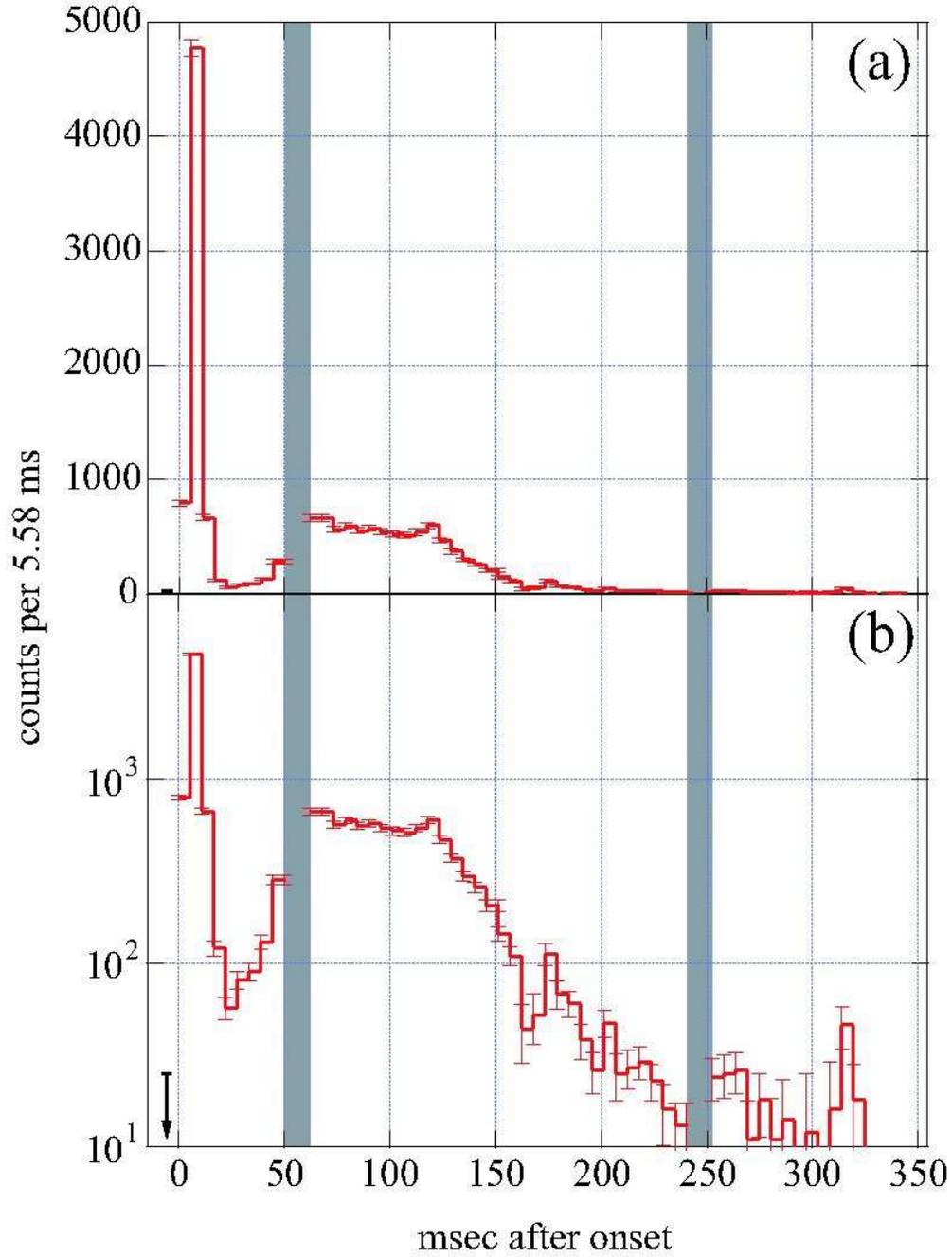}
\caption{The first 350 ms unsaturated peak profile 
of the SGR 1900+14 GF observed with GEOTAIL (a) linear scale, 
and (b) log scale. The time resolution is 5.58 ms
and the energy range is $E\gtrsim 50$ keV. Shaded bars indicate 
the instrumental data gaps of 12 ms.
\label{SGR1900}}
\end{figure}

\clearpage

%% Here we use \plottwo to present two versions of the same figure,
%% one in black and white for print the other in RGB color
%% for online presentation. Note that the caption indicates
%% that a color version of the figure will be available online.
%%

%\begin{figure}
%\epsscale{1.0}
%\plottwo{Fig2_1.eps}{Fig2_2_resize_New.eps}
%\caption{Comparison of the peak profile of the initial spike of 
%SGR giant flare. (a) SGR 1900+14 GF in 1998 (b) SGR 1806-20 GF in 2004.}
%\end{figure}

\begin{figure}
\epsscale{.8}
\plotone{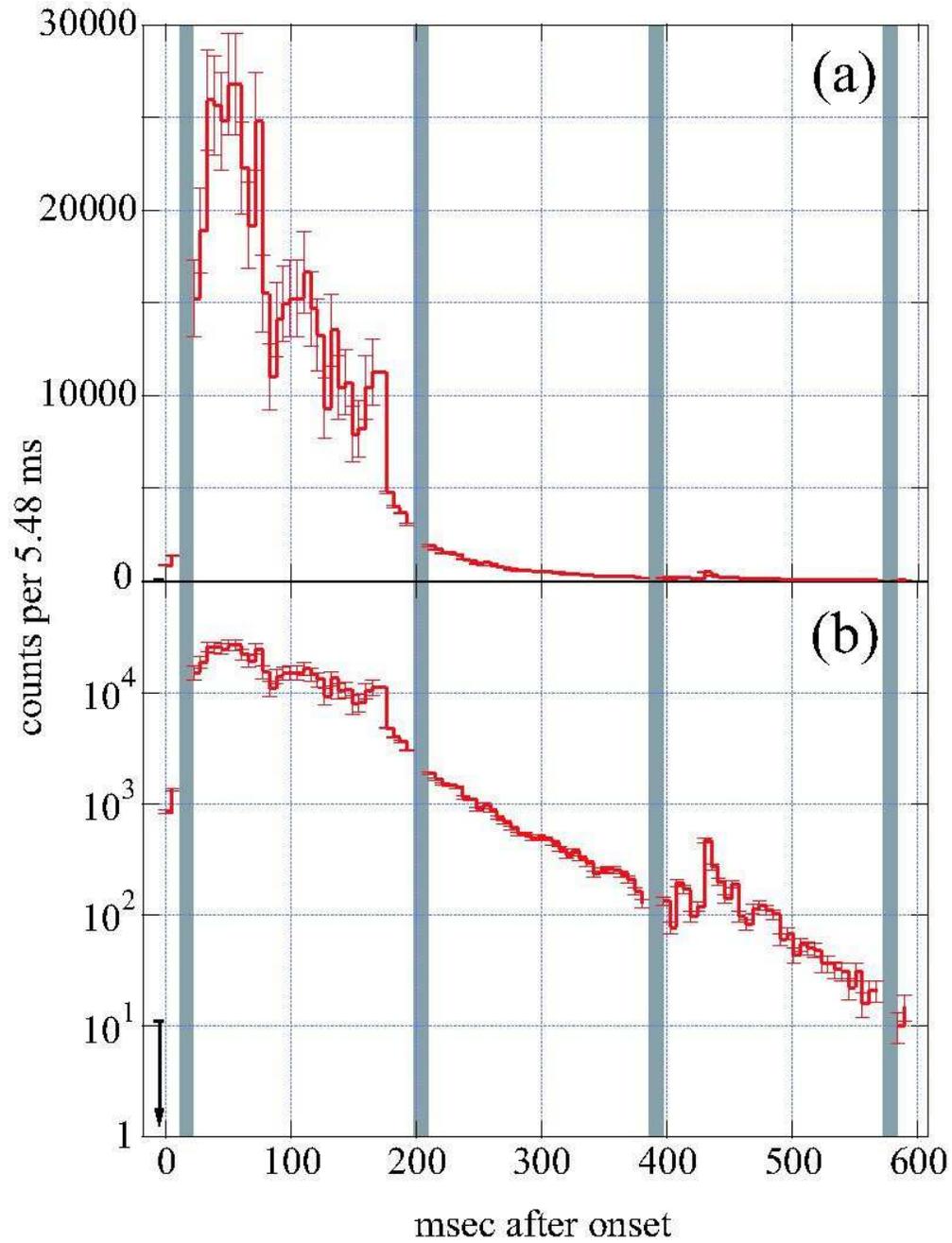}
\caption{
The $E\gtrsim 50$ keV gamma-ray time profile 
of the initial spike of SGR 1806-20 GF 
on 2004 December 27 observed with GEOTAIL (a) linear scale, and 
(b) log scale \citep{terasawa2005}. The time resolution is 5.48 ms.
Shaded bars indicate the instrumental data gaps of 12 ms.
\label{SGR1806}}
\end{figure}

\clearpage

\begin{figure}
%\epsscale{.5}
\plotone{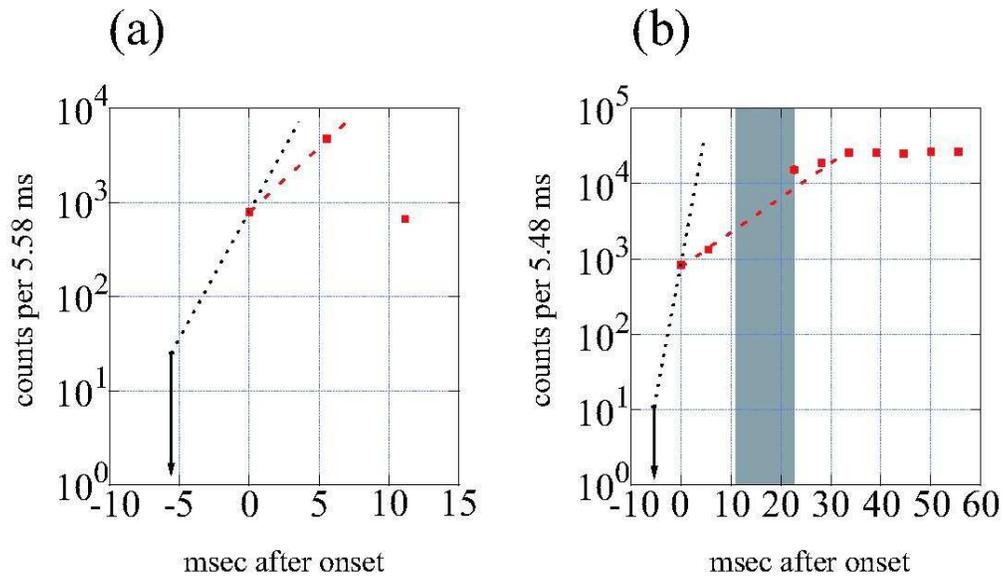}
\caption{Detailed initial rise profiles of the initial spikes
of (a) SGR 1900+14 GF in 1998, and (b) SGR 1806-20 GF in 2004.
The vertical axes are log scale. Two different e-folding rise times 
are clearly seen in both of the initial spikes. The arrow shows the
upper limit of photon counts before the onset. 
\label{Initial}}
\end{figure}

%\clearpage

%\begin{table}
%\begin{center}
%\caption{Estimation of the physical quantities ($E \gtrsim$ 50 keV) 
%of the SGR 1900+14 GF for the two spectral shapes}
%\begin{tabular}{lrr}
%\tableline\tableline
% & OTTB ($kT$=240 keV)\tablenotemark{a} &
%Blackbody ($kT$=230 keV)\tablenotemark{b} \\
%\tableline
% Peak photon flux [photons cm$^{-2}$ s$^{-1}$] & $\left(3.2_{-1.1}^{+4.0}\right)\times10^6$ & $\left(2.4_{-0.9}^{+2.7}\right)\times10^6$ \\
% Peak flux [erg cm$^{-2}$ s$^{-1}$] & $0.85_{-0.30}^{+1.0}$ 
%& $2.4_{-0.9}^{+2.8}$\\
% Peak luminosity [erg s$^{-1}$] & $\left(2.3_{-0.8}^{+2.7}\right)\times10^{46}d_{15}^2$  & $\left(6.4_{-1.2}^{+2.3}\right)\times10^{46} d_{15}^2$ \\
% Fluence [erg cm$^{-2}$] & $\left(1.6_{-0.6}^{+2.0}\right)\times10^{-2}$ &
%$\left(4.5_{-1.6}^{+5.3}\right)\times10^{-2}$ \\
% Total Energy [erg] & $\left(4.3_{-1.5}^{+5.3}\right)\times10^{44} d_{15}^2$  & $\left(1.2_{-0.4}^{+1.4}\right)\times10^{45} d_{15}^2$ \\
%\tableline
%\end{tabular}
%% Any table notes must follow the \end{tabular} command.
%\tablenotetext{a}{\citet{Hurley1999}}
%\tablenotetext{b}{\citet{Boggs2006}}
%\tablecomments{We can also attach a long-ish paragraph of explanatory
%material to a table.}
%\end{center}
%\end{table}

\clearpage

\end{document}